# On the Nonlinear Distortion Characterization in Photovoltaic Modules for Visible Light Communication

Shuyan Chen, *Student Member, IEEE*, Liqiong Liu, *Student Member, IEEE*, and Lian-Kuan Chen, *Senior Member, IEEE*

*Abstract*—Photovoltaic (PV) modules have been employed in visible light communication (VLC) for simultaneous energy harvesting and data reception. A PV-based receiver features easy optical alignment and self-powered operation. It is commonly assumed that PV modules in VLC have a linear optical-electrical response, which is generally true under high illumination levels. This paper will illustrate the exacerbated PV's nonlinear distortion when under typical indoor illumination. The nonlinearity of a PV module for different numbers of PV cells is also characterized. We investigated the transmission performance of a 1-Mbit/s PAM4 signal under different illuminance. Experimental results show that the bit-error rate (BER) first decreases and then increases with the increasing illuminance. Thus, an optimal illuminance to minimize BER exists. In addition, we demonstrated two distortion mitigation methods, namely localized distortion compensation lighting and post-distortion compensation. BER reduction from 3.2×10$^{-1}$ to 2.6×10$^{-3}$ and 2.8×10$^{-2}$ to 1.5×10$^{-2}$ are achieved with the two respective schemes.

## I. Introduction

VISIBLE light communication (VLC) can provide simultaneous data transmission and illumination for various IoT device access [1]. Generally, photodiodes (PD) are used for signal detection at the receiver side. The PD receiver has a very small physical area, thus restricting its detection efficiency for diffused light. With the booming utilization of renewable power, photovoltaic (PV) modules are widely deployed for energy harvesting in many electronic devices. Extending PV's use to VLC applications attracts new research interest. Due to the relatively large physical area, PV modules alleviate the alignment requirement significantly. Prior research focused on designing self-powered VLC receivers with silicon and organic PV modules [2,3], enabling simultaneous communication and energy harvesting.

However, owing to the large physical area of PV module, the junction capacitance is high [4], leading to the low modulation bandwidth. It has been proposed to apply a reverse bias on the PV module to obtain up to 30% bandwidth enhancement [5]. Although the method shows bandwidth improvement, applying reverse bias consumes extra power. Alternatively, the data rate can be increased via multilevel modulation for bandwidth-limited channels [6]. Higher device linearity is required for multilevel modulation [7], which is yet to be investigated for PV modules.

In this paper, the linearity characteristic of a PV module is modeled and experimentally characterized. It is shown that the linearity strongly depends on the illuminance and direct current (DC) bias of the PV module. We also investigate the effect of PV cell numbers on nonlinearity. A multi-cell model is established and verified by experimental results. We conduct a transmission experiment using a PV module-based VLC receiver for a 1-Mbit/s four-level pulse amplitude modulation (PAM4). We further demonstrate two distortion mitigation methods, namely, localized distortion compensation lighting and post-distortion compensation. Reduction of bit-error rate (BER) from 3.2×10$^{-1}$ to 2.6×10$^{-3}$ for a 1- Mbit/s PAM4 transmission is realized using a localized distortion-compensating light source.

## II. Principles

The photovoltaic equation of PV cell's optical-electrical (OE) conversion can be written as [8]:

$$V \approx \frac{nk_BT}{q}\ln\left(\frac{I_{PH}}{I_0} + 1\right) \quad (1)$$

where $V$ is the voltage of the load, $I_{PH}$ is the generated photocurrent, $I_0$ is the reverse saturation current of the diode, $n$ is the diode ideality factor, $k_B$ is Boltzmann's constant, $T$ is the temperature in Kelvin, and $q$ is the electron charge. Eq. (1) shows that the output voltage of the PV module increases logarithmically with $I_{PH}$. The photocurrent $I_{PH}$ is linearly proportional to the light intensity $L$ as $I_{PH} = \eta L$. The $\eta$ is a conversion factor converting illuminance to current, and it depends on the spectral composition of the visible light, physical area, and quantum efficiency of the PV cell. Eq. (1)

Manuscript received xxx xx, 2021; revised xxx xx, 2021; accepted xxx xx, 2021. Date of publication xxx xx, 2021; date of current version xxx xx, 2021. This work was supported in part by the HKSAR RGC grant (GRF 14207220). (Corresponding author: Lian-Kuan Chen.)

Shuyan Chen, Liqiong Liu, and Lian-Kuan Chen are with the Department of Information Engineering, The Chinese University of Hong Kong, Shatin, N. T., Hong Kong SAR. (e-mail: lkchen@ie.cuhk.edu.hk).

Color versions of one or more figures in this letter are available at https://doi.org/xx. xxxx/LPT. xxxx. xxxxxxxx.

Digital Object Identifier xx. xxxx/LTP. xxxx. xxxxxxxx.

also depicts that the PV cell's response *V-I* curve is nonlinear. In practice, a PV module is crafted by connecting different numbers of PV cells in serial to obtain a larger output [8]. Thus, it is also necessary to find the nonlinearity of a PV module as a function of the numbers of connected cells to facilitate spectrally efficient multilevel modulations. A single PV cell under fixed illumination can be modeled as a non-ideal current source, consisting of an ideal current source ($I_{PH}$) and a shunt resistor (*R*). The short-circuit current *I* of a PV module with *N* cells connected in serial can be derived from the total voltage divided by the total internal resistance as:

$$I = \frac{\sum_{i=1}^{N} I_{PH\ i}\ R_i}{\sum_{i=1}^{N} R_i} \quad (2)$$

Under uniform illumination with *N* identical PV cells connected in serial, *R* and $I_{PH}$ will be the same for all cells, thus $I=I_{PH}$. The open-circuit voltage $V_N$ equals $N\times V$, and from Eq. (1), it can be expressed as

$$V_N = N\frac{nk_BT}{q}\ln\left(\frac{\eta L}{I_0}+1\right) \quad (3)$$

It is worth noting that Eq. (1) and (3) hold when the PV module is not reverse biased. If reverse biased, the PV cells work linearly like a photodiode [8]. However, it requires either external bias or the use of self-harvested power for the reverse bias, thus compromising its energy harvesting efficiency. In this paper, we will investigate and quantify the PV module nonlinearity analytically and experimentally. We first evaluate the second-order nonlinearity of the PV module by taking the derivative of Eq. (3) with respect to illuminance *L*.

$$V'_N = \frac{dV_N}{dL} = N\frac{nk_BT}{q\eta L} \quad (4)$$

$$V''_N = \frac{d^2V_N}{dL} = -N\frac{nk_BT}{q\eta L^2} \quad (5)$$

The first-order derivative in Eq. (4) is the slope of OE response curve under different illuminance levels. Eq. (5) characterizes the second-order nonlinearity, and it shows a larger nonlinearity at lower illumination. Other higher-order derivatives can be derived similarly.

## III. EXPERIMENTAL SETUP

Fig. 1 illustrates the experimental setup of the PV-module-based VLC systems. On the transmitter side, random data sequences generated by a computer are first converted to PAM4 signals with a DC bias superimposed by an arbitrary waveform generator (AWG, Siglent SDG 5162). The output electrical signals drive a light-emitting diode (LED, OSRAM LUW W5AM) to generate visible light signals. At the receiver side, a PV module (ANYSOLAR SM301K09L) is connected to a receiving circuit. A resistor load is used to extract the converted electrical power from the PV module. After 1.0-m free-space transmission, the optical signal is detected by the PV module and the voltage on load is recorded by a mixed-signal oscilloscope (MSO, Tektronix MSO 4054) for subsequent offline signal processing. A distortion-compensating LED driven by a DC power is placed beside the PV module to study the effectiveness of nonlinear distortion mitigation. A lux meter (Smart sensor, AS823) is used to estimate the illuminance. Before the experiment, a linear PD is used to measure the linearity of the LED transmitter for reference. The results show that the LED has a linear response within the operation range (3V–10V), thus eliminating the possibility of its contribution to system nonlinearity. For performance evaluation, we estimated the BER performance with a forward-error-correction (FEC) threshold value of $2.0\times 10^{-2}$.

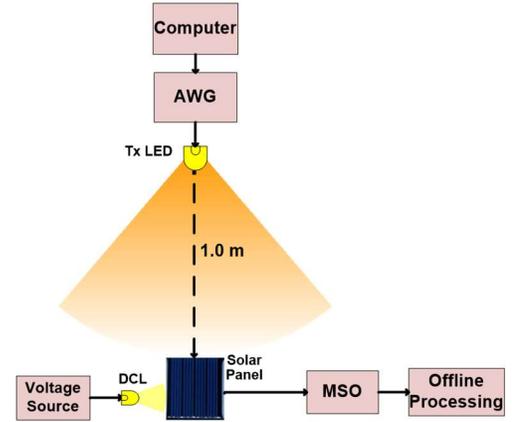

Fig. 1. Experimental setup for solar panel VLC system. AWG: arbitrary waveform generator, DCL: distortion-compensating LED, MSO: mixed-signal oscilloscope.

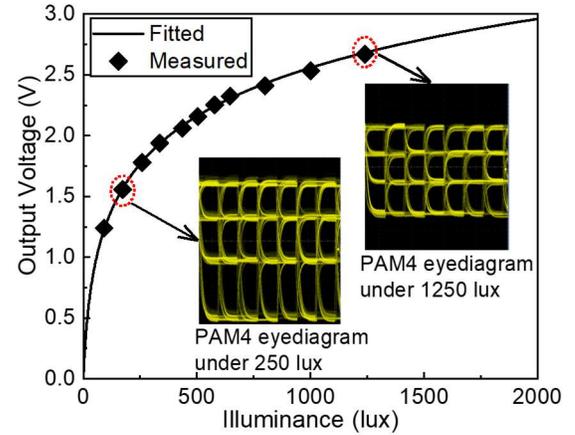

Fig. 2. DC response curve and signal distortion.

## IV. EXPERIMENTAL DEMONSTRATION

In this section, we investigate the nonlinearity of PV receivers experimentally and evaluate the BER performance of PAM4 transmission. Firstly, we characterize the nonlinear OE response of a single-cell PV module. Secondly, the nonlinear response of serially connected multi-cell PV modules is investigated. Moreover, two nonlinear distortion mitigation schemes are demonstrated for PAM4 signals in the PV module-based VLC system.

### A. OE response

We first measured multiple output voltages under the illuminance from 0 to 1000 lux, covering the most indoor lighting scenarios [10]. The data are fitted into Eq. (3) (for *N*=1) to derive the diode ideality factor *n* and the reverse saturation current $I_0$. The measured results and the fitted curve generated by Eq. (3) are illustrated in Fig. 2, showing good agreement between the two. For the illuminance between 0 lux and 1000 lux, the OE response of PV cells exhibits severe nonlinearity, compared to that in the range of 1000 lux to 2000 lux. This is further verified by the two eye diagrams in the insets. The 250-lux eye diagram exhibits significant amplitude suppression at high signal levels, showing severe nonlinear distortion. In contrast, the adjacent level spacings are much more uniform for 1250 lux.

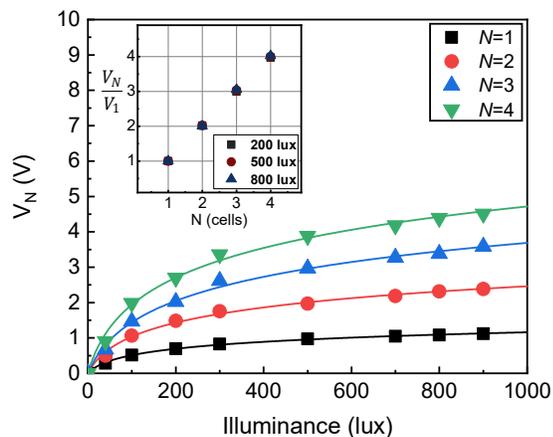

Fig. 3. Measured DC response curve with different numbers of cells connected.

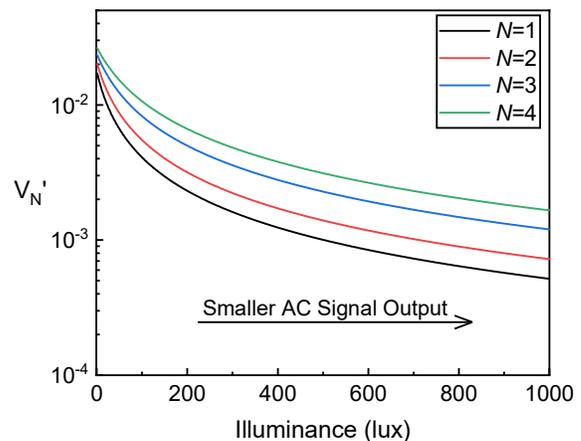

Fig. 4. First-order derivative of solar panels with different numbers of cells.

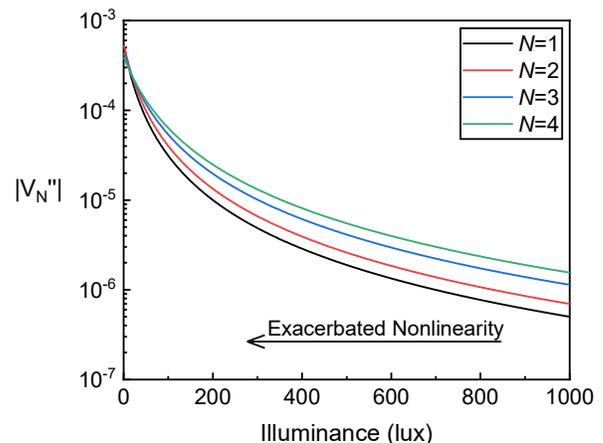

Fig. 5. Second-order derivative of solar panels with different numbers of cells.

*B. OE response of multi-cell PV module*

First, the values of diode ideality factor $n$ and the reverse saturation current $I_0$ for each cell were estimated, and consistent results for different cells were attained. Next, we use Eq. (3) to generate fitted curves for the OE response of multiple PV cells connected in serial, as shown in Fig. 3. The inset illustrates that the output voltage is proportional to the number of serial PV cells for all illuminance under a fixed and uniform illuminance across cells. The slope of the fitted curves at a given illuminance is steeper for PV modules with a larger number of cells, implying a larger alternating current (AC) signal output. By using Eq. (4), Eq. (5), and the estimated $n$ and $I_0$, the derivatives, $V_N'$ and $V_N''$, are plotted in Fig. 4 and Fig. 5, respectively. We can observe that a PV module with more cells gives a higher AC signal response; however, it also exhibits a higher nonlinearity. If we focus on each PV module at low illuminance, the AC signal response is larger, but with exacerbated nonlinearity. The following section will compare the communication performances of two distortion reduction schemes with different illuminance levels.

*C. Nonlinear distortion compensation*

Fig. 6 illustrates the BER of a 1-Mbit/s PAM4 signal received with respect to modulation index under different ambient illuminations. The illuminance is increased by applying a larger DC bias at the transmitter side. The signal amplitude increases with an increased modulation index from 0.05 to 0.3, resulting in BER reduction under all illuminance levels. A further increase in modulation index results in BER degradation due to more considerable nonlinear distortion for the 200-lux and 350-lux cases. Thus, the modulation index should be optimized when working in a nonlinear (low-illuminance) region.

In contrast, the BERs under 500 lux and 650 lux decrease with the increment of the modulation index, indicating that nonlinear distortion is less significant. The results in Fig. 6 also imply that nonlinear distortion caused by PV modules can be mitigated by simply increasing the illuminance. However, the illuminance level may be set for a specific indoor scenario [9] and thus may not be a free parameter. Two distortion reduction schemes are investigated.

We first explored the possibility of using a localized light source near the solar panel to mitigate the nonlinear distortion, but without affecting the set indoor illuminance level. We performed a 1-Mbit/s PAM4 transmission under different illuminations. The illuminance from the transmitter was fixed at 425 lux at the solar panel. The modulation index varied from 0.2 to 0.4. An additional local distortion-compensating LED (DCL) module, as shown in Fig. 1, was placed near the solar panel. Its output power was varied to study its effect on BER and distortion, with the results shown in Fig. 7. For all modulation indexes, an optimum local illuminance that minimizes the BER exists. The initial BER reduction is due to the smaller nonlinear distortion by the increased local illuminance. Then, BER is degraded due to larger shot noises and smaller output signal amplitudes from a flatter slope in the signal response at high illuminance, as shown in Fig. 2.

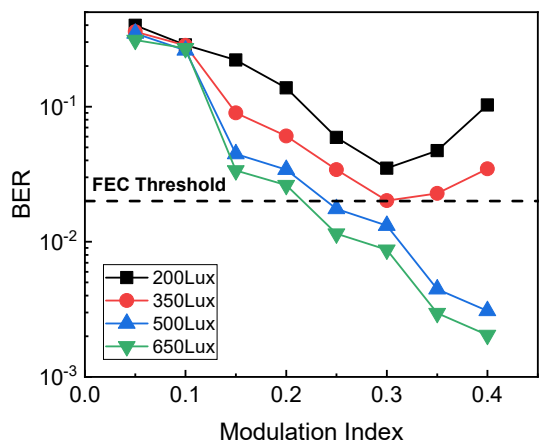

Fig. 6. BER versus modulation index of a 1-Mbit/s PAM4 signal under different illuminance.

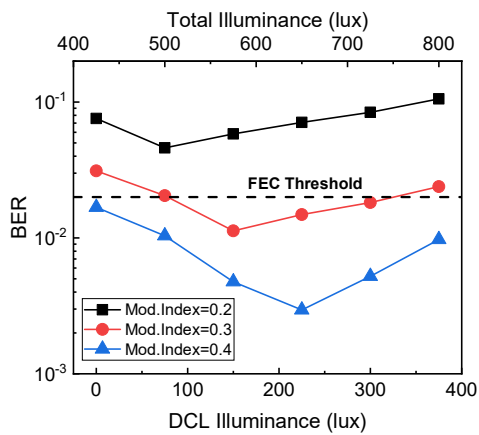

Fig. 7. Transmission of a 1-Mbit/s PAM4 signal under different DCL illuminance for different modulation index.

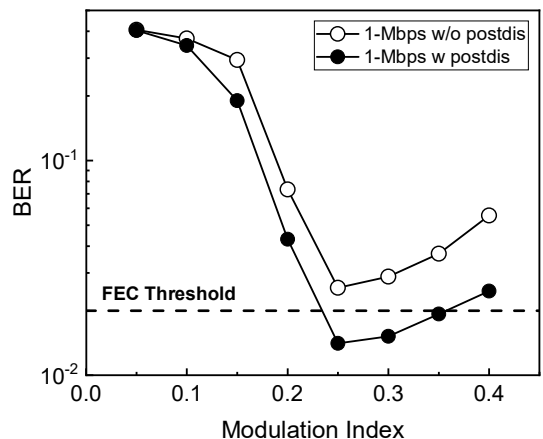

Fig. 8. Transmission of a 1-Mbit/s PAM4 signal under 350 lux with postdistortion nonlinear compensation. Postdis: post-distortion compensation.

We further investigated the distortion reduction by a post-distortion compensation at the receiver using the inverse function of Eq. (3), with the DC component removed first, for offline digital signal processing. This compensation may be applicable for the scenario when indoor illumination adjustment or a local light source is unavailable. As shown in Fig. 8, BER can be reduced for all modulation indexes under 350-lux illuminance by the compensation scheme. Albeit the post-distortion compensation is effective, the increasing trend of BER at high modulation indexes still exists due to imperfect compensation. Imperfect compensation was employed because perfect compensation requires a high amplification, leading to larger amplified noise. If post-distortion digital signal processing is less desirable for some scenarios, real-time analog distortion compensation circuits employing diode array can be used [10].

## V. CONCLUSIONS

In this paper, we showed that PV modules exhibit nonlinear optical-to-electrical conversion in VLC receivers under low illuminance. This is in contrast to prior demonstrations under high illuminance or with reverse-biased PVs, thus with minimal distortion. Low illuminance is more common and suitable for indoor scenarios. The PV's distortion originates from the fact that output voltages increase logarithmically with incident illuminance for PV cells without reverse bias. The nonlinearity of PV modules was analytically studied. We also studied multi-cell PV's distortion and showed that the distortion increases with the number of cascaded cells. Degradation in communication performance was investigated under different illuminance. The reduction of distortion-induced BER degradation by two schemes was demonstrated using a DCL source and post-distortion compensation. A BER reduction from $3.2 \times 10^{-1}$ to $2.6 \times 10^{-3}$ for a 1-Mbit/s PAM4 transmission was realized via DCL.